\documentclass[12pt]{article}
\usepackage{amsmath}
\usepackage{graphicx,psfrag,epsf}
\usepackage{enumerate}
\usepackage{natbib}
\usepackage{url} 
\usepackage{shortcuts}
\usepackage{bbm}

\usepackage{amsthm}
\usepackage{mathtools}
\usepackage{booktabs,multirow}
\usepackage{amsfonts}
\usepackage{subfiles}
\usepackage{float}

\newcommand{\blind}{1}

\let\given\givenbase

\newcommand{\ie}{i.e.}

\numberwithin{equation}{section}
\theoremstyle{plain}
\newtheorem{theorem}{Theorem}[section]

\begin{document}

\def\spacingset#1{\renewcommand{\baselinestretch}%
{#1}\small\normalsize} \spacingset{1}

\date{April 2019}
\if1\blind
{
  \title{\bf More robust estimation of sample average treatment effects using Kernel Optimal Matching in an observational study of spine surgical interventions}
  \author{Nathan Kallus\\
    School of Operations Research and Information Engineering and \\
Cornell Tech, Cornell University, New York, New York 10044\\
    \\
    Brenton Pennicooke \\
    NewYork-Presbyterian Hospital and \\
Weill Cornell Medical Center, New York, New York 10032 \\
\\
    Michele Santacatterina\thanks{
    Corresponding author. This material is based upon work supported by the National Science Foundation under Grants Nos. 1656996 and 1740822.}\hspace{.2cm}\\
    TRIPODS Center for Data Science for Improved Decision Making \\
    and Cornell Tech, Cornell University, New York, New York, 10044}
  \maketitle
} \fi

\if0\blind
{
  \bigskip
  \bigskip
  \bigskip
  \begin{center}
    {\LARGE\bf More robust estimation of sample average treatment effects using Kernel Optimal Matching in an observational study of spine surgical interventions}
\end{center}
  \medskip
} \fi

\newpage
\bigskip
\begin{abstract}
Inverse probability of treatment weighting (IPTW), which has been used to estimate sample average treatment effects (SATE) using observational data, tenuously relies on the positivity assumption and the correct specification of the treatment assignment model, both of which are problematic assumptions in many observational studies. Various methods have been proposed to overcome these challenges, including truncation, covariate-balancing propensity scores, and stable balancing weights. Motivated by an observational study in spine surgery, in which positivity is violated and the true treatment assignment model is unknown, we present the use of optimal balancing by Kernel Optimal Matching (KOM) to estimate SATE. By uniformly controlling the conditional mean squared error of a weighted estimator over a class of models, KOM simultaneously mitigates issues of possible misspecification of the treatment assignment model and is able to handle practical violations of the positivity assumption, as shown in our simulation study. Using data from a clinical registry, we apply KOM to compare two spine surgical interventions and demonstrate how the result matches the conclusions of clinical trials that IPTW estimates spuriously refute. 
\end{abstract}

\noindent%

\vfill

\newpage
\spacingset{1.45} 
\section{Introduction}

Inverse probability of treatment weighting (IPTW) has been used to estimate the sample average treatment effect (SATE) of a treatment on an outcome using observational data. The key idea of IPTW is to correct for selection bias into treatment by weighting each unit in the sample by its probability of being in its treatment group conditional on covariates, \textit{i.e.}, the propensity score \citep{rosenbaum1983central}. In other words, IPTW creates a pseudo-population in which each unit has the same probability of getting treated, thus mimicking a randomized experiment.  IPTW's popular use in medicine \citep{mansournia2016inverse}, epidemiology \citep{hernan2000marginal}, and lately also in computer science \citep{swaminathan2015counterfactual,kallus2018policy, 2018arXiv181102672S} come from its theoretical appeal and interpretability. The standard way to estimate SATE via IPTW consists of predicting the propensity scores by modeling the treatment assignment mechanism, taking their inverse, and plugging the obtained set of weights into a weighted average or a weighted least squares (WLS) estimator \citep{horvitz1952generalization,robins1994estimation,robins2000marginal,lunceford2004stratification}. Wald confidence intervals are then constructed using a robust (sandwich) estimator for the standard error \citep{van2000asymptotic, stefanski2002calculus,freedman2006so,tsiatis2007semiparametric}.  

Positivity, which requires that for any set of covariates it is theoretically possible to observe a unit with either treatment, is key to estimating SATE (without parametric outcome models). However, IPTW's reliance on positivity can be very tenuous. In particular, if positivity is violated in even a very limited region of covariates, then the IPTW estimator for SATE has \textit{infinite} variance \citep{robins2000marginal,cole2008constructing}. Even if positivity holds theoretically, if some propensities are close to 0, then even small errors in propensity estimates can lead to outsize errors in IPTW's SATE estimate. This issue is known as \textit{practical violations of the positivity assumption} \citep{petersen2012diagnosing} and it is well known that it can lead to extreme weights and large variance \citep{robins1995analysis,scharfstein1999adjusting,robins2007comment,kang2007demystifying}, which pose serious problems in practice.

IPTW also relies on the correct specification of the unknown treatment assignment model -- a concern in almost every observational study.

One example of practical positivity violation and possible model misspecification that we study in this paper is in the evaluation of laminectomy alone compared with fusion-plus-laminectomy in patients with lumbar stenosis and lumbar spondylolisthesis. The comparison is based purely on passive observations of historical spine surgical interventions and their outcomes, as recorded in a clinical registry of spine surgeries. Lumbar stenosis is a spine pathology consisting of a compression of the lower back's nerves. Lumbar spondylolisthesis is a pathology in which one vertebra move out of position. Common spine surgical practice suggests treating patients with lumbar stenosis with laminectomy alone, while those with lumbar spondylolisthesis with fusion-plus-laminectomy. While deviations exist, this leads to very limited positivity in the data. Understanding the differing benefits of these treatments is of utmost interest because of the invasive nature of the surgeries. Registry data provide a unique opportunity to use a large number of observations to study these effects, but very limited positivity and potential misspecification remain a significant hurdle to the use of standard methodologies.

Several statistical methods have been proposed to overcome issues of practical positivity violation and potential misspecification. To control for practical positivity violation, the most popular solution is truncation, which consists of replacing outlying weights with less extreme ones \citep{cole2008constructing}. \citet{kang2016practice} and \citet{lee2011weight} investigated the impact of different cutoff points in the distribution of the propensity scores with respect to bias and efficiency. \citet{cole2008constructing} suggested truncating at high percentiles of the distribution of the estimated weights, \textit{e.g.}, the 1st and 99th percentiles. \citet{ju2017adaptive} proposed an adaptive truncation method based on the collaborative targeted maximum likelihood estimation methodology. Despite the fact that truncation reduces the variance of the weights and consequently that of the weighted estimator, it can also introduce substantial bias.  Rather than truncating, \citet{santacatterina2018optimal} and \citet{santa2018} proposed to find the closest set of weights to the IPTW weights while controlling precision by constraining the variance of the resulting estimator or the variance of the weights.

To mitigate the effect of possible misspecification of the treatment assignment model, \citet{imai2014covariate} proposed to use the Covariate Balancing Propensity Score (CBPS), which, instead of plugging in logistic-regression
estimates of propensities, uses IPTW with propensities predicted by the logistic model that balances covariates, found via the generalized method of moments. 
\citet{lee2010improving} proposed to use boosted classification and regression trees to estimate the propensities. \citet{zubizarreta2015stable} proposed Stable Balancing Weights (SBW), which are the set of weights of minimal sample variance that satisfy a list of approximate moment matching conditions to a level of balance specified by the research.

In this paper, we use Kernel Optimal Matching (KOM), a subclass of the Generalized Optimal Matching (GOM) framework \citep{kallus2016generalized}, to provide weights that simultaneously mitigate the effects of possible misspecification of the treatment assignment model and control for possible practical positivity violations. 
We do so by minimizing the worst-case conditional mean squared error of the weighted estimator in estimating SATE over the space of weights. Specifically, KOM controls for practical positivity violations by limiting the variance of the estimate (either by penalizing or constraining it), while mitigating possible model misspecification by using flexible models to balance covariates.
To use KOM, we show how to extend the general approach of \citet{kallus2016generalized}, which focused only on SATT for simplicity, to the case of SATE, which requires a new, more intricate error decomposition and an approach that balances both the conditional average of the control and the treatment outcomes.
Compared with the state-of-the-art methods, we find that estimating SATE with KOM has the advantages of (1) optimally balancing covariates while simultaneously controlling for precision, (2) mitigating the effect of possible misspecification of the treatment assignment model, (3) controlling for strong practical positivity violations, (4) tractably allowing for nonlinear and nonadditive covariates relationships by using kernels, (5) better handling of many covariates and higher order relationships, and (6) automatic selection of balancing levels. In particular, in the simulation study presented in Section \ref{Simu}, we show that both bias and mean squared error of the KOM estimates of SATE are lower than those obtained by using IPTW, truncated IPTW, Propensity Score Matching (PSM), Regression Adjustment (RA), CBPS, and SBW in most of the considered scenarios (we provide a detailed file containing the \textsf{R} code to compute KOM as supporting material). Motivated by this, we use KOM to address the problem of estimating the effect of spine surgical interventions using clinical registry data and find that, whereas both an unadjusted comparison and IPTW show a large significant effect, our estimates show a small insignificant effect, which actually matches the results of recent clinical trials \citep{ghogawala2016laminectomy,forsth2016randomized}.

In the next Section, we introduce a study on the effect of two spine surgical interventions among patients with lumbar stenosis or lumbar spondylolisthesis  that motivated the use of KOM. In Section \ref{kom} we introduce KOM for SATE and discuss practical guidelines (Section \ref{guide}). In Section \ref{Simu}, we present the results of a simulation study aimed at comparing KOM with IPTW, truncated IPTW, PSM, RA, CBPS, and SBW. In Section \ref{qod_cs} we apply KOM to estimate the effect of laminectomy alone versus fusion-plus-laminectomy on the Oswestry Disability Index (ODI) among patients with lumbar stenosis or lumbar spondylolisthesis. We conclude with some remarks in Section \ref{conc}.

\section{The effect of two spine surgical interventions among patients with lumbar stenosis or lumbar spondylolisthesis}
\label{intro_qod}

Lumbar stenosis is a pathology caused by the narrowing of the central spinal canal by overgrown and inflamed connective tissue \citep{resnick2014guideline}. Lumbar spondylolisthesis is a pathology caused by the slippage of one vertebra on another. These spinal pathologies can severely restrict function, walking ability, and quality of life \citep{waterman2012low}. If the symptoms due to lumbar stenosis or lumbar spondylolisthesis are no longer controlled by medications, physical therapy, or spinal injections, then surgery may be needed to improve a patient's symptoms \citep{waterman2012low}. Typically, a laminectomy alone is done to treat lumbar stenosis and a fusion-plus-laminectomy is done to treat lumbar spondylolistheses \citep{resnick2014guideline, eck2014guideline,raad2018trends}.  In addition, patients with leg pain are typically treated with a laminectomy alone, while patients with mechanical back pain are treated with fusion-plus-laminectomy \citep{resnick2014guideline}. Though there is some variation and both interventions may be used for both pathologies, the prevalence of this surgical practice leads to a practical positivity violation when evaluating the effect of laminectomy alone versus fusion-plus-laminectomy in observational data. In particular, in the case study presented in Section \ref{qod_cs}, in which we compare these two spine surgical interventions, less than 10\% of patients with lumbar spondylolisthesis were treated with laminectomy alone and only 1\% of those with a moderate-low leg pain were treated with fusion-plus-laminectomy. 

Due to practical and methodological challenges, randomized trials on the effect of spine surgical interventions are rare \citep{cook2009challenges}. Consequently, most assessments of spine surgical interventions must be based on observational data, in which the true treatment assignment mechanism is hardly ever known and the true causal parameter is hidden by confouding factors. The patient's principal spine pathology, \textit{i.e.}, lumbar stenosis or lumbar spondylolisthesis, is one example of such a confounding factor in this case. Patients with lumbar stenosis, who are mainly treated with laminectomy alone, are also more likely to have a lower ODI overall \citep{pearson2011predominant}. Given these challenges, it is therefore of paramount importance to develop and use statistical methods that provide robust estimates of the SATE for spine surgical interventions based on observational data.

\section{Kernel Optimal Matching}
\label{kom}

In this Section we propose to use KOM for estimating SATE to address the issues noted above. KOM is an approach that minimizes an estimation error objective when unknown conditional expectations are let to vary in a Reproducing Kernel Hilbert Spaces (RKHS) \citep{kallus2016generalized}. To extend this approach to SATE estimation: we analyze the conditional mean squared error (CMSE) of any weighted estimator for SATE; show that the CMSE can be decomposed in terms of the discrepancies in the conditional expectations of the two potential outcomes as well as a variance term and some additional ignorable terms; embed these conditional expectations in an RKHS to obtain an error objective that can be evaluated given observational data; and finally minimize this objective using quadratic optimization to find optimal weights.
We discuss how to automatically tune the method in order to appropriately set the level of balance, the exchange between balance and variance, and the kernel parameters.

\subsection{Decomposing the CMSE for SATE}
\label{deco_cmse}

Suppose we have a simple random sample with replacement of size $n$ from a population. Using the potential outcome framework \citep{imbens2015causal}, for each unit $i=1, \ldots, n$, we let $Y_i(t)$ be the potential outcome of treatment $t\in\lbrace0,1\rbrace$, $X_i$ the observed confounders, $T_i$ the observed treatment, and $Y_i=Y_i(T_i)$ the observed outcome. 
Let $X_{1:n}$ and $T_{1:n}$ denote all the observed confounders and treatment assignments. We impose the assumptions of consistency, non-interference, and ignorability \citep{imbens2015causal}. The assumptions of consistency and non-interference (also known as the SUTVA assumption) state that the observed outcome corresponds to the potential outcome under the treatment applied to that specific unit, \textit{i.e.}, $Y_i = Y_i(t)$, and that the potential outcomes are well-defined. The assumption of ignorability states that the potential outcomes are independent to the treatment assignment once we condition on observed covariates. 
In other words, ignorability states that we have collected all potential confounders in our covariates.  
It suffices to impose the independence in expectation, \textit{i.e.}, we assume only that $\E[Y_i(t)\mid X_i,T_i]=\E[Y_i(t)\mid X_i]$ for $t=0,1$.

We consider estimating the SATE, defined as
\begin{equation}
\begin{aligned}
\op{SATE} &=\dfrac{1}{n}\sum_{i = 1}^n  (Y_i(1)-Y_i(0)),
\end{aligned}
\end{equation}
by using the weighted estimator
\begin{equation}
\begin{aligned}
	\hat{\tau}^{\op{SATE}}_{W} &= \sum_{i:T_i=1}W_iY_i-\sum_{i:T_i=0}W_iY_i=\sum_{i = 1}^n W_i (2T_i-1)Y_i,
\end{aligned}
\end{equation}
which compares the reweighted average outcome among the treated and control group.
Given any weights $W_{1:n}$, setting $W'_i=W_i/\sum_{j:T_j=T_i}W_i$, we have that $\hat{\tau}^{\op{SATE}}_{W'}$ is equivalent to the WLS estimator with weights $W_{1:n}$. In particular, if $\sum_{i:T_i=1}W_i=\sum_{i:T_i=0}W_i=1$ then $W'_{1:n}=W_{1:n}$ and $\hat{\tau}^{\op{SATE}}_{W}$ is already the WLS estimator.

If we were to let $W_i=T_i/e(X_i)+(1-T_i)/(1-e(X_i))$, where $e(X_i) = \mathbb{P}(T_i=1 \given X_i)$ is the propensity score, then $\hat{\tau}^{\op{SATE}}_{\text{W}}$ reduces to the well-known IPTW estimator \citep{horvitz1952generalization,robins1994estimation,lunceford2004stratification}. Similarly, if we normalize these weights to sum to one in each treatment group, then $\hat{\tau}^{\op{SATE}}_{\text{W}}$ reduces to the WLS estimator with IPTW weights.
Instead of taking this plug-in approach, we will find the weights $W_{1:n}$ that optimize an error objective given by the CMSE.

We now decompose the error of the weighted estimator $\hat{\tau}^{\op{SATE}}_{\text{W}}$ for \textit{any} weights $W_{1:n}$ that are a function of the covariate and treatment data, $X_{1:n},T_{1:n}$, \ie~ $W_i = W(X_{1:n},T_{1:n})$. We start by defining $f_t(X_i)=\Efb{Y_i(t)\mid X_i}$ and $\sigma^2_{i}=\op{Var}\prns{Y_i\mid X_i,T_i}$. Next define the conditional average of SATE (CSATE):
$$
\op{CSATE}=\Efb{\op{SATE}\mid X_{1:n}}=\frac1n\sum_{i=1}^n(f_1(X_i)-f_0(X_i)).
$$
In our decomposition, we will separate out the error of the weighted estimator in estimating just CSATE, which is what we will actually focus on. 

For any function $f$,
we define the $f$-moment discrepancy between the weighted $t$-treated group and the whole sample as
\begin{align}
B_t(W_{1:n};f)&=\sum_{i=1}^n\prns{\indic{T_i=t}W_i-\frac1n}f(X_i),
\label{bias0}
\end{align}
where $\mathbbm{I}[T_i=t]\in\{0,1\}$ is the indicator for unit $i$ having treatment $t$. We now decompose the conditional bias and CMSE of $\hat{\tau}^{\op{SATE}}_{\text{W}}$. 

\begin{theorem}
\label{th1} Suppose $W_{1:n}$ is independent of all else given $X_{1:n},T_{1:n}$. Then, under consistency, non-interference, and ignorability, 
\begin{align}
	\mathbbm{E}\left[\hat{\tau}^{\op{SATE}}_{\text{W}} - \op{SATE} \given  X_{1:n}, T_{1:n} \right] &= \mathbbm{E}\left[\hat{\tau}^{\op{SATE}}_{\text{W}} - \op{CSATE} \given  X_{1:n}, T_{1:n} \right] \label{thmbias} \\&=  B_1(W_{1:n};f_1) - B_0(W_{1:n};f_0)
	\notag\\
	\mathbbm{E}\left[\prns{\hat{\tau}^{\op{SATE}}_{\text{W}} - \op{CSATE}}^2 \given  X_{1:n}, T_{1:n} \right] &= 
	(B_1(W_{1:n};f_1) - B_0(W_{1:n};f_0))^2 + \sum_{i = 1}^n W_i^2 \sigma_i^2 \label{thmcmse_csate}
	\\
	\mathbbm{E}\left[\prns{\hat{\tau}^{\op{SATE}}_{\text{W}} - \op{SATE}}^2 \given  X_{1:n}, T_{1:n} \right] &= 
(B_1(W_{1:n};f_1) - B_0(W_{1:n};f_0))^2 + \sum_{i = 1}^n W_i^2 \sigma_i^2
	\label{thmcmse_sate}\\ &\phantom{=}+ \frac1{n^2}\sum_{i=1}^n\op{Var}(Y_i(1)-Y_i(0)\mid X_i) \notag\\
	\notag
	&\phantom{=}+
	\frac1n\sum_{i=1}^nW_i(2T_i-1)\op{Cov}(Y_i,Y_i(1)-Y_i(0)\mid X_i,T_i).
\end{align}
\end{theorem}

\noindent
Theorem \ref{th1} shows that the bias of $\hat{\tau}^{\op{SATE}}_{W}$ decomposes into two discrepancies: the discrepancy in the $f_1$ moment between the weighted treated group and the whole sample and the discrepancy in the $f_0$ moment between the weighted control group and the whole sample (eq.~\eqref{thmbias}). 
Next, Theorem \ref{th1} shows that the CMSE of $\hat{\tau}^{\op{SATE}}_{W}$ in estimating CSATE decomposes into a conditional bias squared plus a conditional variance, where the conditional variance is simply given by the weighted squared Euclidean norm of the $W$ vector, with components weighted appropriately by the conditional variance of the outcomes (eq.~\eqref{thmcmse_csate}). This allows us to understand precisely where errors due to the choice of $W_{1:n}$ arise from and help us in judicially choosing $W_{1:n}$ to minimize total error. In particular, we will next discuss an approach to minimize this total error, given some restrictions on the unknown $f_0,f_1$.

Theorem \ref{th1} also shows that the CMSE of $\hat{\tau}^{\op{SATE}}_{W}$ in estimating SATE differs from that of estimating CSATE by two certain terms. We next argue it is safe to ignore these terms when using this CMSE objective to choose $W_{1:n}$.
One additional term (the second on the right-hand side of eq.~\eqref{thmcmse_sate}) corresponds to the variance of SATE in estimating CSATE (or, vice versa). In particular, this term is both small and \textit{independent} of $W_{1:n}$, so it should not affect how we choose $W_{1:n}$ and we may ignore it. Another additional term (the third on the right-hand side of eq.~\eqref{thmcmse_sate}) 
involves both the weights $W_i$ and the covariance of the observed outcome ($Y_i$) and the individual effect ($Y_i(1)-Y_i(0)$). Although this term does involve the weights, it is always small for any set of weights. In particular, if conditional variances are bounded such that $\op{Var}(Y_i(t)\mid X_i)\leq\sigma^2_{\max}$ (as would be the case under homoskedasticity, for example) and if we focus our attention to weights that sum to one in each treatment group (as we do in this paper) then, applying the Cauchy-Schwarz inequality to the covariance and H\"older's inequality to the sum, we see that this term is bounded by $4\sigma^2_{\max}/n$ regardless of the choice of $W_{1:n}$. Otherwise, using the Cauchy-Schwarz inequality to bound the unknowable conditional covariance of the two potential outcomes by their respective conditional variances \citep[see also][]{splawa1990application,imai2008variance}, we simply get an additional term that we could easily also take into consideration if we so choose.

\subsection{Worst-case squared bias}
\label{nsqcb}

The bias of the weighted estimator, and correspondingly its CMSE, depends on the unknown functions $f_1,f_0$.
In this Section, we propose to minimize the worst-case CMSE and correspondingly replace the bias by its worst-case value, normalized relative to the ``size'' of $f_1,f_0$ since the bias scales linearly in these functions.

To define this ``size,'' we embed each function in a normed space. In particular, we consider an extended seminorm $\| \cdot \|_t$, \ie, a norm on functions from the space of covariates to the space of outcomes that can also assign the values $0$ and $\infty$ to nonzero elements.
We then define the ``size'' of the pair $f_1,f_0$ as $\sqrt{\fmagd {f_1}_1^2+\fmagd {f_0}_0^2}$ (\ie, we take the product of the spaces). 
Given this magnitude (we discuss our specific choice below), we can define the relative worst-case squared bias as follows:
\begin{equation}
\label{nbias}
\mathcal{B}(W_{1:n})=  \sup_{f_0,f_1}\frac{B_1(W_{1:n};f_1)-B_0(W_{1:n};f_0)}{\sqrt{\|f_{1}\|_1^2+ \|f_{0}\|_0^2}}
= \sqrt{\Delta^2_1(W_{1:n}) + \Delta^2_0(W_{1:n})},
\end{equation}
where 
$$
\Delta_t(W_{1:n})=\sup_{f} \frac{B_t(W_{1:n};f)}{\|f\|_{t}}=\sup_{\|f\|_{t} \leq 1} B_t(W_{1:n};f)
$$
is the relative worst-case discrepancy in the $f$ moment between the weighted $t$-treated group and the whole sample over all $f$ functions in the unit ball of $\|\cdot\|_t$.

In particular, we will use the norm given by an RKHS. Given a positive semidefinite (PSD) kernel $\mathcal K_t(x,x')$, these norms take the form
$$
\fmagd f_t = \inf\braces{
{{\displaystyle\sum_{i,j=1}^\infty}\alpha_i\alpha_j\mathcal K_t(x_i,x_j)}
:
f=\sum_{i=1}^\infty\alpha_i\mathcal K_t(x_i,\cdot),\,
\sum_{i=1}^\infty\alpha_i^2\mathcal K_t(x_i,x_i)<\infty
}.
$$
Despite this complex form of the norm, the corresponding form for $\Delta_t(W)$ is rather simple.
Define the matrix $K_t\in\mathbb R^{n\times n}$ as $K_{tij}=\mathcal K_t(X_{i},X_{j})$ (that such a matrix is PSD for any set of points is precisely the definition of a PSD kernel). Then, we have that
\begin{align*}
\Delta^2_t(W_{1:n}) &=  \sup_{\|f\|_{t} \leq 1} \prns{\sum_{i = 1}^n  \prns{W_i \mathbbm{1}[T_i=t] - \dfrac{1}{n} } f_t(X_i)}^2 \\
&= \sup_{\sum_{i,j=1}^n\alpha_i\alpha_j\mathcal K_t(X_i,X_j)\leq1} \prns{\sum_{i = 1}^n  \prns{W_i \mathbbm{1}[T_i=t] - \dfrac{1}{n} } \sum_{j=1}^n\alpha_j\mathcal K_t(X_i,X_j)}^2\\
&= \sup_{\alpha^TK_t\alpha\leq1} \prns{\alpha^TK_t(I_tW_{1:n}-e_n)}^2\\
&=(I_{t}W_{1:n}-e_n)^TK_t(I_{t}W_{1:n}-e_n) \\
&=W_{1:n}^TI_{t}K_tI_{t}W_{1:n}-2e_n^TK_tI_{t}W_{1:n}+e_n^TK_te_n,
\end{align*}
where $e_n$ is the length-$n$ vector with $1/n$ in every entry and $I_t$ is the $n$-by-$n$ diagonal matrix with $\indic{T_i=t}$ in the $i\thh$ diagonal entry. The second equality above follows by the representer theorem, which states that when optimizing over an RKHS norm ball it is sufficient to restrict to span of the kernels at the points where the function is evaluated \citep{scholkopf2001learning}. The third equality follows by rewriting using matrix notation. The fourth equality follows by basic Euclidean geometry and the fifth by expanding the matrix product.

\subsection{Minimizing the worst-case CMSE}
\label{cmse}

In the previous two Sections we decomposed the conditional mean squared error and defined the relative worst-case squared bias. 
If we also estimate (or, bound) the conditional variances $\sigma_i^2$,
this immediately leads to an objective for the worst-case CMSE. We propose to estimate SATE using the weighted estimator with weights the minimize the worst-case CMSE of this estimator. We restrict to weights that sum to one in each treatment group, which is equivalent to just using the WLS estimator for any given unrestricted nonnegative weights.
Formally, we let $\mathcal W=\fbraces{W_{1:n}\in\R n:W_i\geq0\;\forall i,\,\sum_{i:T_i=1}W_i=\sum_{i:T_i=0}W_i=1}$ and
choose the weights $W_{1:n}$ to solve the optimization problem
\begin{align}
\notag
&\underset{W_{1:n} \in \mathcal{W}}{\min}\sup_{\|f_{1}\|_1^2+ \|f_{0}\|_0^2\leq1}
\mathbbm{E}\left[\prns{\hat{\tau}^{\op{SATE}}_{\text{W}} - \op{CSATE}}^2\mid X_{1:n},T_{1:n}\right]
\\&\qquad\qquad\qquad\qquad\qquad\qquad=\underset{W_{1:n} \in \mathcal{W}}{\min}
\prns{\Delta^2_1(W_{1:n})+\Delta^2_0(W_{1:n})+\sum_{i = 1}^n W_i^2 \sigma_i^2}
.\label{kom_cmse}
\end{align}
By minimizing the worst-case CMSE, this optimization problem essentially finds weights that optimally balance the confounders (by minimizing the relative worst-case discrepancies) while simultaneously controlling precision (by regularizing the norm of $W_{1:n}$). In particular, the worst-case discrepancies $\Delta_t(W_{1:n})$ are precisely a distributional distance (specifically, an integral probability metric) between the sample distribution of covariates and the reweighted $t$-treated-group distribution of covariates.

If we use an RKHS norm as we have in the last section then this optimization problem reduces to a linearly-constrained convex-quadratic optimization problem:
\begin{equation}
\label{kom1}
\underset{\substack{W_{1:n}\geq0,\\ W_{1:n}^TI_1e_n=W_{1:n}^TI_0e_n=n}}{\min}
W_{1:n}^T(I_{1}K_1I_{1}+I_{0}K_0I_{0}+\Sigma)W_{1:n}
-2e_n^T(K_1I_{1}+K_0I_{0})W_{1:n},
\end{equation}
where $\Sigma$ is the $n$-by-$n$ diagonal matrix with $\sigma_i^2$ in its $i\thh$ diagonal entry. This optimization problem can be easily and quickly solved by many off-the-shelf solvers (in particular, the problem can be efficiently solved by a polynomial-time algorithm). We use Gurobi \citep{gurobi}, for example.

\section{Practical guidelines for choosing kernels and conditional variances}
\label{guide}

In the previous Sections we formulated a novel KOM approach to find optimal weights for estimating SATE. This, however, depended on a choice of kernel and conditional variances.
Indeed, the solutions to the optimization problem \eqref{kom1} depends on these choices.

We generally propose to use a polynomial Mahalanobis kernel:
\begin{equation}\label{polykernel}
\mathcal K_t(x,x')=\gamma_t(1+\theta_t(x-\hat\mu_n)^T\hat\Sigma_n^{-1}(x'-\hat\mu_n))^d,
\end{equation}
where $\hat\mu_n$ is the sample mean of confounders and $\hat\Sigma_n$ their sample covariance (in other word, we simply Studentize the data first).
This kernel has a few hyperparameters: $\gamma_t$, $\theta_t$, and $d$.
The parameter $d$ controls the degree of the polynomial kernel. We generally suggest to use $2$ or $3$ mostly based on the numerical results from simulations in the following Section. This choice for $d$ offers the model some flexiblity to balance higher order moments of covariates, while the other hyperparameters allow us to control the relative importance of such higher orders. In particular, KOM with polynomial kernel degree 3 outperforms IPTW, truncated IPTW, PSM, RA, CBPS and SBW with respect to both bias and MSE across all levels of practical positivity violation in our simulations in Section~\ref{Simu}.

We suggest to choose the other two hyperparameters, $\gamma_t$ and $\theta_t$, as well as the conditional variance parameters, $\sigma_i^2$, in a data-driven way.
The parameter $\theta_t$ controls the relative importance of higher-order moments: a lower value stresses more balance in lower-order moments over higher-order moments. We would like to chose this to match the level of nonlinearity of $f_t$.
Finally, the parameter $\gamma_t$ controls the overall scale of the kernel and we would like to chose it to match the scales of $f_t$.
In particular, to achieve this, we suggest to tune $\gamma_t$ and $\theta_t$ using the empirical Bayes approach of marginal likelihood \citep{rasmussen2004gaussian}. Specifically, we suppose $f_1,f_0$ came from a Gaussian process with kernels $\mathcal K_1,\mathcal K_0$ and that each $Y_i$ was observed from $f_{T_i}(X_i)$ with Gaussian noise of variance $\lambda_{T_i}^2$. We then choose the values for $\gamma_t$, $\theta_t$, $\lambda_t$ that maximize the likelihood of the data and we set $\sigma_i^2=\lambda_{T_i}^2$. This has various unique benefits, such as automatically learning the structure of the data and preferring simpler models by default. This method is also implemented in the \textsf{matlab} package \textsf{GPML}. In our code, we provide a sufficient re-implementation in \textsf{R}.

Of course, there are many other possible choices and one of the benefits of the KOM approach is its great flexibility. For example, one may use the Gaussian or Mat\'ern kernels \citep{scholkopf2001learning} instead of the polynomial, or even much more complicated kernels \citep{wilson2013gaussian}. Additionally, instead of Studentizing the data, we could instead parameterize the matrix in the inner product used in the polynomial, Gaussian, or Mat\'ern kernel (\ie, replace $\hat\Sigma_n^{-1}$ in eq.~\eqref{polykernel} by a parameter matrix $\Omega$) and learn that matrix as part of the marginal likelihood tuning step. For example, an approach known as Automatic Relevance Detection (ARD) is to use a diagonal matrix with tunable variable-specific weights on the diagonal. This allows us to learn the importance of different variables and appropriately stress the balance in the different variables and their interactions.

\section{Simulations}
\label{Simu}

In this Section, we compare the performance of KOM with IPTW, truncated IPTW (tIPTW), Propensity Score Matching (PSM), Regression Adjustment (RA), CBPS and SBW with respect to bias and MSE in estimating SATE in various linear, nonlinear, correct, and misspecified scenarios and across different levels of strength of practical violation of the positivity assumption.
All bias and MSE values are computed over 500 replications. 

\subsection{Setup}
\label{Simu_setup}

We considered a sample size of $n=200$. For the linear scenario we drew data from the following model: $Y_i = \alpha + \delta T_i + \sum_{k=1}^K X_{i,k} + N(0,1)$, where  $T_i \sim \text{binom}(\pi_i)$, $\pi_i = \text{expit}(\beta(\sum_{k=1}^K X_{i,k}))$, $X_{k,i} \sim \text{N}(0,1)$, $k=1,\dots,K$, and $K=2$. For the nonlinear scenario, we drew data from the following model: $Y_i = \alpha_1 + \delta T_i + \sum_{k=1}^K X_{i,k} + \sum_{k=1}^K X_{i,k}^2 + \sum_{k \neq m} X_{i,k}X_{i,m} + N(0,1)$, where  $T_i \sim \text{binom}(\pi_i)$, $\pi_i = \text{expit}(\beta( \sum_{k=1}^K X_{i,k} + \sum_{k=1}^K X_{i,k}^2 + \sum_{k \neq m} X_{i,k}X_{i,m} ))$, $X_{k,i} \sim \text{N}(0,1)$, $k=1,\dots,K$, and $K=2$. The intercepts $\alpha$ and $\alpha_1$ were chosen so that the marginal mean of $Y_i$ was equal to 0. We set the true causal parameter $\delta=1$. 
We vary $\beta$ in order to vary the level of practical positivity violation.

In particular, we considered seven equally-spaced values, ranging from 0.1 to 3, for the $\beta$ parameter in the treatment assignment models above. By tuning this parameter, we can easily control the strength of practical positivity violation, where higher values correspond to a strong practical positivity violation. For instance, in the linear scenario, under the weakest level considered ($\beta=0.1$), the propensity score ranged on average between 0.5 and 0.8, while under the strongest level ($\beta=3$) between 0.002 and 0.999 (average of min/max over replications).

For the correct scenarios we plugged into the models the correct variables, $X_1$ and $X_2$. We refer to these scenarios as correct. To evaluate the performance under misspecification, we also generated $Z_1 = (2+X_1)/(\exp (X_1))$ and $Z_2 = ((X_1X_2/25)+1)^3$ and plugged them into the models instead of the correct variables $X_1$ and $X_2$. We refer to these scenarios as misspecified.

For each scenario and in each sample, we then computed the set of KOM, IPTW, tIPTW, PSM, CBPS and SBW weights. Specifically, under the linear correct scenario, we computed the set of KOM weights by using a linear kernel (KOM-$\mathcal{K}_1$), IPTW and PSM weights by regressing the treatment on the linear terms using logistic regression, and CBPS and SBW weights by including the linear terms in the covariates fed to the methods. We refer to the last four as linear IPTW, PSM, CBPS, and SBW. Under the nonlinear correct scenario, we computed the set of KOM weights by using a polynomial degree 2 kernel (KOM-$\mathcal{K}_2$), IPTW and PSM weights by regressing the treatment on the linear, quadratic and interaction terms using logistic regression, and CBPS and SBW weights by including linear, quadratic and interaction terms in the covariates fed to the methods. We refer to the last four as quadratic IPTW, PSM, CBPS, and SBW. Under both linear and nonlinear misspecified scenarios, we computed the set of KOM weights by using a polynomial degree 3 kernel (KOM-$\mathcal{K}_3$), IPTW and PSM weights by regressing the treatment on the linear, quadratic, cubic and interaction terms (all monomials up to degree three) using logistic regression, and CBPS and SBW weights by including all monomials up to degree three in the covariates fed to the methods. We refer to the last four as cubic IPTW, PSM, CBPS and SBW. We specified the level of balance for SBW to be equal to 1/100 \citep{zubizarreta2015stable}. If SBW failed to find a solution, we increased the level of balance to 1/10 and then to 1 if that also failed. For each scenario and each level of the strength of practical positivity violation we also computed a set of truncated IPTW weights. Specifically, we truncated the IPTW weights at the 1st and 99th percentile of their distribution as suggested by \citet{cole2008constructing}. To compute the KOM weights, we rescaled the covariates to have mean 0 and variance 1 and tuned the hyperparameters by using Gaussian process marginal likelihood, as described in our practical guidelines in Section \ref{guide}. 

Given a set of weights,
we estimated the SATE by using a WLS estimator, regressing the outcome on the treatment, weighted by KOM, IPTW, tIPTW, PSM,  CBPS, and SBW. To estimate SATE via RA, we computed, for each scenario and each level of practical positivity violation, the contrasts of means of treatment-specific predicted outcomes. We used the \textsf{R} interface of \textsf{Gurobi} to obtain the set of KOM weights, and the \textsf{glm}, \textsf{CBPS} and \textsf{sbw} packages to obtain the set of IPTW, tIPTW, CBPS and SBW weights respectively. We also chose \textsf{Gurobi} as solver to obtain the SBW weights.  We used the \textsf{R} package \textsf{Matching} with the default settings \citep{sekhon2011multivariate} to perform PSM. We used \textsf{lm} for RA. 

\subsection{Results}

In this section we present and discuss the simulations results obtained across levels of practical positivity violation in the correct linear and nonlinear scenarios (Section \ref{clnls}), in the misspecified linear scenario (Section \ref{mls}) and in the misspecified nonlinear scenario (Section \ref{mnls}). In summary, KOM outperformed IPTW, tIPTW, PSM, RA, CBPS and SBW with respect to bias and MSE across all levels of practical positivity violation and considered scenarios. In addition, KOM outperformed the other methods especially under strong practical positivity violation.

\subsubsection{Correct linear and nonlinear scenarios}
\label{clnls}

\begin{figure}[H] 
\begin{center}
\includegraphics[scale=.65]{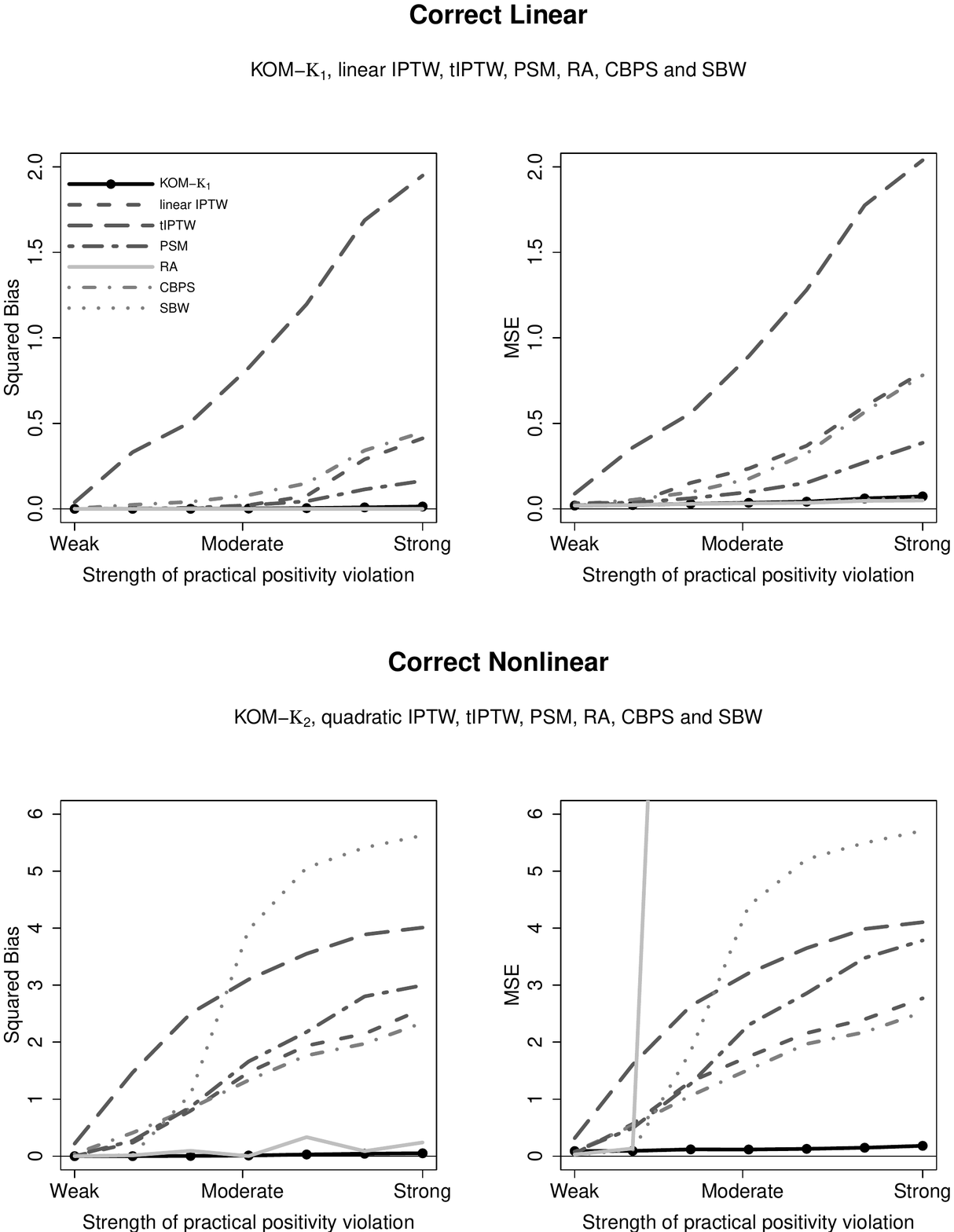}
\end{center}
\caption{\footnotesize Squared bias (left panels) and MSE (right panels) of the estimated SATE using KOM (solid-circle), IPTW (dashed), tIPTW (long-dashed), PSM (two-dashed), RA (solid), CBPS (dashed-dotted) and SBW (dotted) when increasing the strength of practical positivity violation in the correct linear scenario (top panels) and in the correct nonlinear scenario (bottom panels), $n=200$. Top panels shows the results when using KOM-$\mathcal{K}_1$, linear IPTW, tIPTW, PSM, RA, CBPS, and SBW. Bottom panels show the results when using KOM-$\mathcal{K}_2$, quadratic IPTW, tIPTW, PSM, RA, CBPS, and SBW.
\label{fig1} }
\end{figure}

Figure \ref{fig1} shows squared bias (left panels) and MSE (right panels) of KOM (solid-circle), IPTW (dashed), tIPTW (long-dashed), PSM (two-dashed), RA (solid), CBPS (dot-dashed) and SBW (dotted) in the correct linear scenario (top panels) and correct nonlinear scenario (bottom panels). Under the correct linear scenario, KOM-$\mathcal{K}_1$ outperformed IPTW, tIPTW, PSM, and CBPS with respect to both bias and MSE. It is worth mentioning that, while, the bias and the MSE of IPTW, tIPTW, PSM, and CBPS increased with the levels of practical positivity violation, those of KOM-$\mathcal{K}_1$ were consistently low across all levels. Notably, in the linear scenarios, linear SBW and KOM-$\mathcal{K}_1$ performed similarly since both control a similar linear moment discrepancy of just a few (two) covariates. KOM-$\mathcal{K}_1$ also performed similarly to RA. In Section \ref{comparison} we show that KOM, which optimizes these discrepancies directly, outperforms SBW with respect to both bias and MSE, and it outperforms RA with respect to MSE, in these linear scenarios when the number of confounders considered is increased. In the nonlinear correct, misspecified linear, and misspecified nonlinear scenarios, KOM also outperformed SBW and RA, as discussed below.

The lower panels of Figure \ref{fig1} show the bias and the MSE across levels of practical positivity violation under the correct nonlinear scenario. KOM-$\mathcal{K}_2$ outperformed IPTW, tIPTW, PSM, CBPS and SBW with respect of both bias and MSE across all considered levels of practical positivity violation. It is worth mentioning that the bias of KOM-$\mathcal{K}_2$ was as low as that of the RA, which is theoretical zero given the fact that RA used the correct model. In addition, contrary to RA, KOM-$\mathcal{K}_2$ also resulted in a low MSE while that of RA exploded when increasing the level of practical positivity violation. Although KOM and RA can be thought as methodologically similar techniques, the results of our simulation study suggest that KOM-$\mathcal{K}_2$ can be used even in nonlinear settings without being affected by moderate or strong practical positivity violation.   

\subsubsection{Misspecified linear scenario}
\label{mls}

The top panels of Figure \ref{fig2} shows squared bias (left panels) and MSE (right panels) of KOM (solid-circle), IPTW (dashed), tIPTW (long-dashed), PSM (two-dashed), RA (solid), CBPS (dashed-dotted) and SBW (dotted) in the misspecified linear scenario. 
In this scenario, we observed that the cubic variants of methods better handle the misspecification compared with the linear ones. We therefore focus only on the results obtained from KOM-$\mathcal{K}_3$, cubic IPTW, tIPTW, PSM, RA, CBPS, and SBW.
KOM-$\mathcal{K}_3$, a polynomial degree 3 kernel for KOM, outperformed cubic IPTW, tIPTW, PSM, RA, CBPS, and SBW  across all considered levels of practical positivity violation. Cubic RA resulted in very high bias and MSE across all levels (results are outside the plot region in Fig. \ref{fig2}).

\subsubsection{Misspecified nonlinear scenario}
\label{mnls}

\begin{figure}[H] 
\begin{center}
\includegraphics[scale=.65]{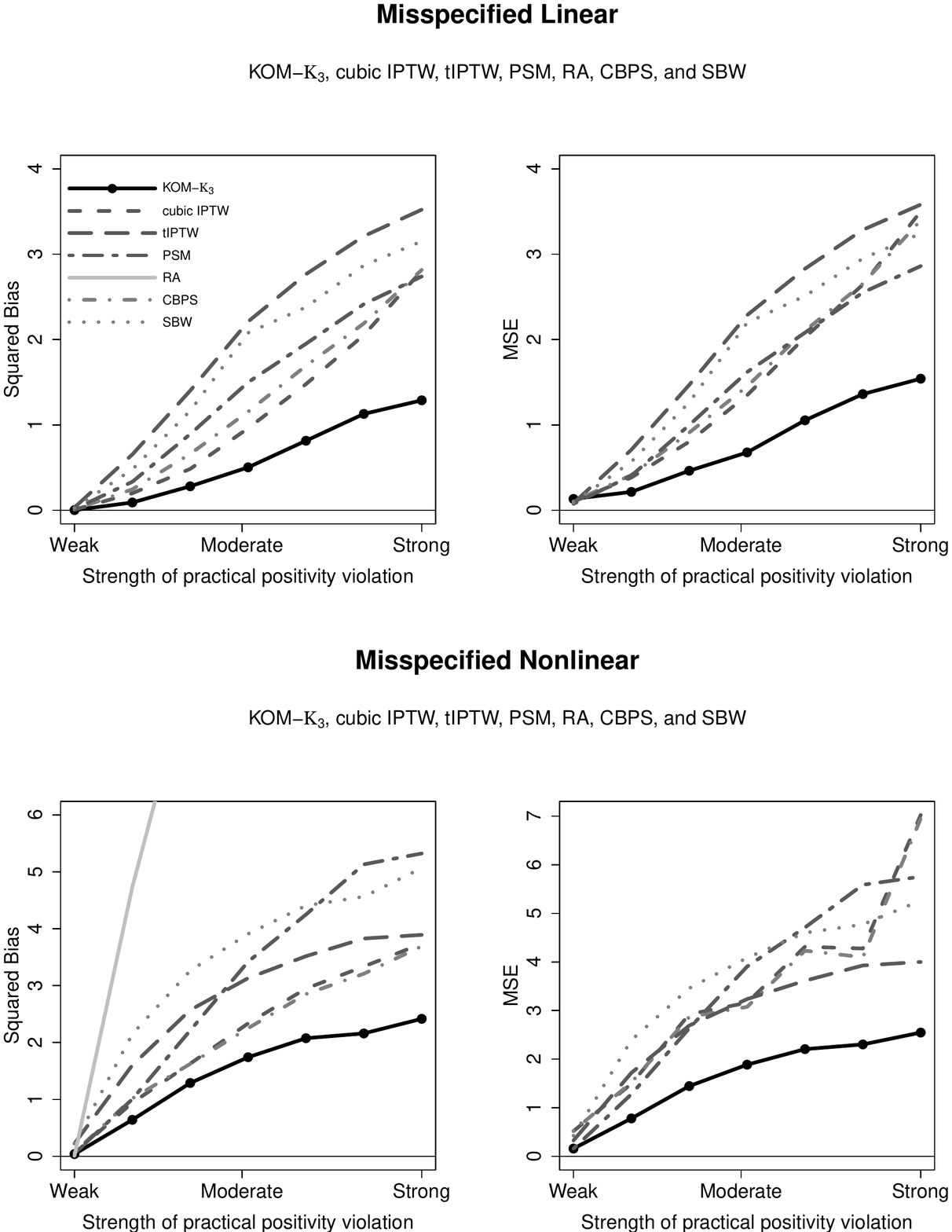}
\end{center}
\caption{\footnotesize Squared bias (left panels) and MSE (right panels) of the estimated SATE using KOM-$\mathcal{K}_3$ (solid-circle), cubic IPTW (dashed), cubic tIPTW (long-dashed), cubic PSM (two-dashed), cubic RA (solid; outside of plot region in 3 of 4 plots), cubic CBPS (dashed-dotted), and cubic SBW (dotted)  when increasing the strength of practical positivity violation in the misspecified linear scenario (top panels) and in the misspecified nonlinear scenario (bottom panels), $n=200$. 
\label{fig2} }
\end{figure}

The bottom panels of Figure \ref{fig2} shows squared bias (left panel) and MSE (right panel) of KOM (solid-circle), IPTW (dashed), tIPTW (long-dashed), PSM (two-dashed), RA (solid), CBPS (dashed-dotted) and SBW (dotted) in the misspecified nonlinear scenario. 
KOM-$\mathcal{K}_3$ outperformed cubic IPTW, tIPTW, PSM, RA, CBPS, and SBW. Cubic RA resulted in very high bias and MSE across all levels of practical positivity violation (results are outside the plot region in Fig. \ref{fig2}).

In summary, KOM showed a consistently lower bias and MSE across all considered scenarios and across levels of practical positivity violation, and especially under strong practical violation. These results suggest that KOM with a polynomial degree $d \geq 2$ kernel mitigates the impact of model misspecification while being able to handle strong practical positivity violations.

\subsubsection{Linear SBW, RA and KOM-$\mathcal K_1$ when increasing the number of confounders under linear scenarios}
\label{comparison}
The results presented in the top panels of Figure \ref{fig1} 
suggest that in the correct linear scenario 
when the number of confounders considered was equal to 2, KOM performed similarly to SBW and RA with respect to bias and MSE. Motivated by the fact that in practice (including in our own application), the number of confounders used for analysis can be much larger, in this Section we present the results of a simulation study aimed at comparing bias and MSE of KOM, SBW and RA when increasing the number of confounders.  Specifically, we drew data from the following model: $Y_i = \alpha + \delta T_i + \sum_{k=1}^K X_{i,k} + N(0,1)$, where  $T_i \sim \text{binom}(\pi_i)$, $\pi_i = \text{expit}(\beta(\sum_{k=1}^K X_{i,k}))$, $\delta=1$, and $X_{k,i} \sim \text{N}(0,1), k=1,\dots,K$, with $K=2,20,50$ and $100$. We set $\beta=2$ for a moderately strong practical positivity violation 
and computed bias and MSE over 500 replications in the correct linear scenario with a sample size of $n=200$. 

\begin{figure}[H] 
\begin{center}
\includegraphics[scale=.65]{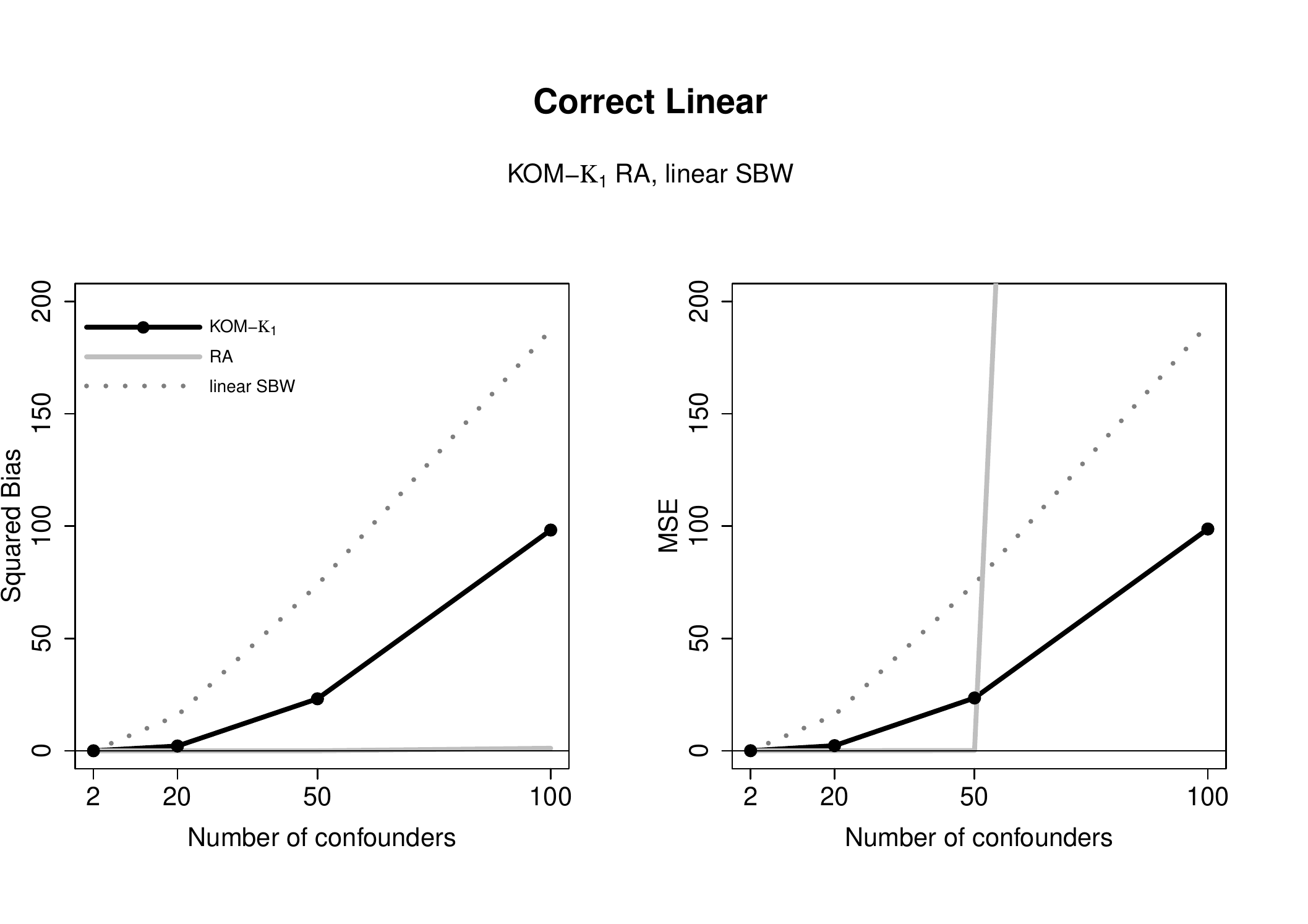}
\end{center}
\caption{\footnotesize Squared bias (left panels) and MSE (right panels) of the estimated SATE using KOM-$\mathcal{K}_1$ (solid-circle), RA (solid) and liner SBW (dotted) when increasing the number of confounders (2, 20, 50, 100) in the linear correct scenario, $n=200$.
\label{fig4} }
\end{figure}

Figure \ref{fig4} shows squared bias (left panels) and MSE (right panels) in the correct linear scenario across $K=2,20,50$, and $100$ number of confounders. KOM outperformed SBW with respect to bias and MSE across all considered numbers of confounders, suggesting that KOM provides lower bias and MSE compared with SBW when the number of confounders is moderate. KOM outperformed RA with respect to MSE when the number of confounders increased.

\subsection{Coverage}

To compute confidence intervals of a weighted estimator for SATE, Wald confidence intervals can be used together with the robust sandwich estimator \citep{hernan2001marginal,robins2000marginal,freedman2006so}. 
We next compare the empirical coverage of such 95$\%$ confidence intervals for the various methods across scenarios under the strongest practical positivity violation setting.
In the case of PSM, we use the standard error estimator proposed by \citet{abadie2006large}. Table \ref{table1} shows the results. In summary, KOM achieved desirable coverage under both linear and nonlinear correct scenarios. These results are similar to those found by \cite{kallus2016generalized}[Section 4.4] in which coverage was computed keeping $X_{1:n}$ and $T_{1:n}$ fixed. Since all methods had significant bias in the misspecified scenarios, they all exhibit undercoverage, as expected. The slightly higher coverage of IPTW, PSM and CBPS with cubic logistic models simply arises from their much larger variance, leading to very wide confidence intervals. Indeed, when truncating the IPTW weights, leading to lower variance without affecting bias (see bottom panels of Fig.~\ref{fig2}), the coverage drops to 0$\%$.

\begin{table}
\centering
\caption{Empirical coverage of Wald 95\% confidence intervals \label{table1}}
\begin{tabular}{lllllll}
\toprule
                      \multirow{2}{*}{\textbf{Scenario}} & \multicolumn{6}{c}{\textbf{Method}}                                                 \\ 
                       & KOM  & IPTW            & tIPTW           & SBW             & CBPS            & PSM  \\ \midrule 
Correct linear         & 0.92 & 0.45            & 0.05            & 0.95            & 0.45            & 0.88 \\
Correct nonlinear      & 0.88 & \textless{}0.01 & \textless{}0.01 & \textless{}0.01 & \textless{}0.01 & 0.69 \\
Misspecified linear      & 0.27 & \textless{}0.01 & \textless{}0.01 & \textless{}0.01 & \textless{}0.01 & 0.73 \\
Misspecified nonlinear & 0.02 & 0.16            & \textless{}0.01 & 0.01            & 0.09            & 0.10 \\ \bottomrule
\end{tabular}
\end{table}

\subsection{Computational time of KOM}

In this Section we report the computational time required by KOM in the simulation study described in Section \ref{Simu_setup}. Three steps are required to compute the set of KOM weights. First, we tune the kernel's hyperparameters; second, we construct the matrices required by problem \eqref{kom1}; and third we solve problem \eqref{kom1}. We computed the computational time by using the \textsf{R} package \textbf{rbenchmark} on a AWS EC2 C5 instance, Intel Xeon Platinum 8000 series, 3.5 GHz, 16GB RAM and a Linux Ubuntu 16.04 operating system.

In the correct linear scenario with $n=200$, KOM required a mean computational time of 2 seconds to obtain the weights. Tuning the hyperparameters required 50$\%$  of the computational time, computing the matrices 49$\%$, and solving the optimization problem 1$\%$. Similar mean computational times were observed in the misspecified linear scenario and in the correct nonlinear scenario.  In the misspecified nonlinear scenario, KOM required a mean computational time of 3.2 seconds to obtain the weights. Tuning the hyperparameters required 70$\%$  of the computational time, computing the matrices 29$\%$, and solving the optimization problem 1$\%$. The mean computational times were similar across levels of practical positivity violation.

\section{Application to the study of spine surgical interventions}
\label{qod_cs}

In this Section we apply KOM to the observational study presented in Section \ref{intro_qod}. We used data from a single-institutional subset of the Spine QOD registry \citep{qod}. The registry was launched in 2012 with the aim of evaluating the effectiveness of spine surgery interventions on the improvement of pain, disability, and quality of life. QOD contains clinical and demographic information as well as patient-reported outcomes.  We evaluated the effect of fusion-plus-laminectomy compared to laminectomy alone on the Oswestry Disability Index (ODI), an index used by surgeons to quantify disability, for the treatment of lumbar stenosis or spondylolisthesis. Previous randomized control trials have shown that fusion-plus-laminectomy and laminectomy alone have equivalent average improvement on the ODI of patients with these conditions \citep{ghogawala2016laminectomy,forsth2016randomized}.

\subsection{Study population and models setup}

We restrict our study to \textit{primary surgery}, defined as the first spine surgery intervention for each patient. Patients were interviewed before surgical intervention, and  demographic and clinical information was collected. ODI was collected at 3-month follow-up. The study subset was composed of 311 patients, 247 of which received laminectomy alone and 64 fusion-plus-laminectomy. As described in Section \ref{intro_qod}, spine surgical practice may lead to a practical violation of the positivity assumption. In our dataset, $1\%$ of those patients with a moderate-low leg pain were treated with fusion-plus-laminectomy and less than $10\%$ of the patients with lumbar spondylolistheses were treated with laminectomy alone.

We identified as potential confounders the following variables: lumbar stenosis (yes vs. no), lumbar spondylolistheses (yes vs. no), leg pain (score from 0 to 10), back pain (score from 0 to 10), activity outside home (yes vs. no), activity at home (yes vs. no), duration of symptom (less than 3 months vs. greater than or equal to 3 months), motor deficiency (yes vs. no), dominant symptoms (back; leg; both), and age at interview. Common statistical practice suggest using IPTW to consistently estimate the effect of laminectomy alone versus fusion-plus-laminectomy in the presence of these confounders.
To apply IPTW, we used logistic regression to estimate the propensities and compute the set of IPTW weights by taking their inverse. Based on the simulation results showed in Section \ref{mls} and \ref{mnls}, we used a cubic logistic regression models (IPTW$_3$). We also compute the set of KOM weights by using a polynomial kernel degree 3 (KOM-$\mathcal{K}_3$). We tuned the kernel's hyperparameters using Gaussian process marginal likelihood and solve problem \eqref{kom1} by using quadratic optimization.  

We considered the following model to evaluate the effect of fusion-plus-laminectomy $(T=1)$ versus laminectomy alone $(T=0)$ on ODI among patients with lumbar stenosis or spondylolisthesis,
\begin{equation}
\label{wols}
 	\mathbbm{E}[Y_i(1)] = \beta_1 + \beta_2 \mathbbm{I}[T=1],
\end{equation}
%
where $\mathbbm{I}[T=1]$ is the indicator function for fusion-plus-laminectomy, $Y_i(T)$ is the potential outcome of observing ODI under intervention $T$ for the \textit{i}-th unit, $\beta_1$ is the effect of laminectomy alone and $\beta_2$ is the SATE. We estimated $\beta_2$
using either ordinary least squares (unweighted) or weighted least squares with weights given either by IPTW$_3$ or KOM-$\mathcal{K}_3$. We computed robust (sandwich) standard errors in each case. We used the \textsf{R} interface of \textsf{Gurobi} to obtain the set of KOM weights, and the \textsf{glm} package and the \textsf{poly} function to obtain the set of IPTW weights. We used the \textsf{R} package \textsf{sandwich} to estimate robust standard errors.


\begin{table}
\centering
\caption{The effect of fusion-plus-laminectomy on ODI \label{tab1}}
\setlength{\tabcolsep}{1em}
\begin{tabular}{llll}\toprule
& Naive      & IPTW$_3$ &  KOM-$\mathcal{K}_3$  \\
\midrule
$\hat{\beta}_2$ (SE) & 5.1* (2.3) &  9.7* (4.6)          & 0.5 (4.4)       \\\bottomrule\\[-1em]
\multicolumn{4}{c}{* indicates statistical significance at the 0.05 level.}
\end{tabular}
\end{table}


\subsection{Results}

Table \ref{tab1} shows the results of our analysis.
When analysing the distribution of IPTW$_3$ weights, a weight of more than 1,000 was assigned to $n=28$ patients, suggesting a strong practical positivity violation. Both the naive estimator ($\hat \beta_2=5.1$; SE: $2.3$) and IPTW$_3$ ($\hat \beta_2=9.7$; SE: $4.6$) indicated a statistically significant positive effect of fusion-plus-laminectomy compared with laminectomy alone on ODI. In contrast, and similar to the results obtained by two recent randomized controlled trials \citep{ghogawala2016laminectomy,forsth2016randomized}, KOM-$\mathcal{K}_3$ resulted in an estimated effect that is both much smaller in magnitude and is statistically insignificant ($\beta_2=0.5$; SE: $4.0$). Whereas an analysis based on IPTW leads to conclusions that perhaps spuriously refute experimental evidence, using KOM we conclude, in agreement with experimental evidence, that among patients with lumbar stenosis or spondylolisthesis, fusion-plus-laminectomy did not result in better ODI compared with laminectomy alone.

\subsubsection{Results when changing model degree}
The results of the simulation study presented in Section \ref{Simu} suggested that the cubic variants of all considered methods better handled model misspecification compared with the linear ones. This led us to use cubic variants in estimating the effect of fusion-plus-laminectomy compared with laminectomy alone on ODI in the above.
In this Section we study the change in these estimates if we change the degree, $d$, of the polynomial models considered in KOM and IPTW. Specifically, we let the degree of the polynomials range from 1, corresponding to the linear kernel for KOM and a linear logistic regression model for IPTW, to 5, corresponding to a polynomial kernel of degree 5 for KOM and a quintic logistic regression model for IPTW. The results are shown in Table \ref{tab2}.

IPTW (second row of Table \ref{tab2}) led to volatile estimates that switched back and forth in both sign and significance as we varied the degree. In contrast, KOM led to stable results that decreased in magnitude from a narrowly significant effect, similar to that of the naive estimator, to a statistically insignificant effect, similar to that of the experimental results, as we increased the degree (first row of Table \ref{tab2}). These results suggest first that KOM results in more stable estimates and that using KOM with a nonlinear kernel ($d\geq2$) leads to improved control of confounders and consequently to more coherent clinical results.


\begin{table}
\centering
\caption{Effect estimates when increasing the degree of polynomials\label{tab2}}
\setlength{\tabcolsep}{1em}
    \begin{tabular}{llllll}\toprule
        $\hat{\beta}_2$ (SE) & Linear      & Quadratic &  Cubic & Quartic & Quintic  \\
        \midrule
        KOM & 4.6* (2.3) &    2.1 (2.8)   & 0.5 (4.4) & 1.5 (4.6) & 0.7 (4.8)\\
        IPTW & 2.0 (3.0) & -3.3 (4.2)  & 9.7* (4.6) & 7.6* (3.3) & 4.5 (3.7)
        \\\bottomrule\\[-1em]
\multicolumn{4}{c}{* indicates statistical significance at the 0.05 level.}
\end{tabular}
\end{table}


\section{Conclusions}
\label{conc}

In this paper, we developed an approach using KOM to provide weights for the estimation of SATE. The method developed directly and optimally controls the total error --- both bias and variance -- of the estimates uniformly over a class of models given by a RKHS. This leads the method to effectively mitigate issues of possible misspecification and robustly handle moderate and strong practical positivity violations, two issues that are of central concern in many observational studies.

By using mathematical optimization, KOM optimally minimizes the conditional mean squared error of any weighted estimator with respect to the weights, resulting in a lower bias and MSE compared with IPTW, tIPTW, PSM, RA, CBPS and SBW {in most of} the considered scenarios of our simulation study. In addition, KOM automatically learns the structure of the data and allows the researcher to balance linear, nonlinear, additive, and non-additive covariate relationships without sacrificing performance.

Alternative formulations of the optimization problem \eqref{kom1} can be used. For instance, we may limit precision by bounding the variance of the resulting weighted estimator up to a level specified by the researcher rather than regularizing it. Additionally, we may impose different norms on the conditional expectation functions of potential outcomes and even constrain them to be equal up to a constant shift or separately regularize their difference (effect) and their average (baseline). These may provide improvements in certain settings where such structure holds.




\bibliographystyle{chicago}
\bibliography{bib}

\end{document}